\documentclass[%
reprint,
superscriptaddress,
 amsmath,amssymb,
 aps,
prl,
longbibliography,
floatfix,
]{revtex4-2}

\usepackage[dvipsnames,svgnames,x11names,hyperref]{xcolor}
\usepackage{siunitx}
\usepackage{tabularx}
\usepackage{color}
\usepackage{graphicx}
\usepackage{dcolumn}
\usepackage{bm}
\usepackage{hyperref}
\usepackage[mathlines]{lineno}
\usepackage{mathrsfs}
\usepackage{braket}
\usepackage{placeins} 
\usepackage{float} 
\usepackage{tikz}

\newcommand{\hide}[1]{}

\newcommand{\fig}[1]{Fig.\,\ref{#1}}

\hypersetup{colorlinks=true, 
	linkcolor={blue!75!black!80!yellow},
	citecolor={blue!75!black!80!yellow}, 
	urlcolor={blue!75!black!80!yellow}
	}
\makeatletter \renewcommand\@make@capt@title[2]{%
\@ifx@empty\float@link{\@firstofone}{\expandafter\href\expandafter{\float@link}}%
\sffamily{\textbf{#1}}\@caption@fignum@sep#2 }

\usepackage[normalem]{ulem}

\begin{document}

\title{Cavity-modified unimolecular dissociation reactions \\ \textit{via} intramolecular vibrational energy redistribution}

\author{Derek S. Wang}
\email{derekwang@g.harvard.edu}
\affiliation{Harvard John A. Paulson School of Engineering and Applied Sciences, Harvard University, Cambridge, MA 02138, USA}

\author{Tom\'{a}\v{s} Neuman}
\affiliation{IPCMS de Strasbourg, UMR 7504 (CNRS – Universit\'e de Strasbourg), 67034 Strasbourg, France}

\author{Susanne F. Yelin}
\email{syelin@physics.harvard.edu}
\affiliation{Department of Physics, Harvard University, Cambridge, MA 02138, USA}

\author{Johannes Flick}
\email{jflick@flatironinstitute.org}
\affiliation{Center for Computational Quantum Physics, Flatiron Institute, New York, NY 10010, USA}

\begin{abstract}
\noindent While the emerging field of vibrational polariton chemistry has the potential to overcome traditional limitations of synthetic chemistry, the underlying mechanism is not yet well understood. Here, we explore how the dynamics of unimolecular dissociation reactions that are rate-limited by intramolecular vibrational energy redistribution (IVR) can be modified inside an infrared optical cavity. We study a classical model of a bent triatomic molecule, where the two outer atoms are bound by anharmonic Morse potentials to the center atom coupled to a harmonic bending mode. We show that an optical cavity resonantly coupled to particular anharmonic vibrational modes can interfere with IVR and alter unimolecular dissociation reaction rates when the cavity mode acts as a reservoir for vibrational energy. We find a strong dependence on the initial state of the cavity and molecule. In particular, when the cavity is initially empty, the dissociation rate decreases, while when the cavity is initially hotter than the molecule, the cavity can instead accelerate the reaction rate. These results lay the foundation for further theoretical work toward understanding the intriguing experimental results of vibrational polaritonic chemistry within the context of IVR.
\end{abstract}
\date{\today}

\maketitle

\section{Introduction} \label{sec:intro}
Mode-selective chemistry---exciting just a single bond in a molecule to control its chemical properties---is a long sought-after goal that would allow selective control over chemical reactivity and energy transfer \cite{Frei1980, Frei1983, Frei1983a, Bucksbaum1990, brumer1986control, Warren1581}. Realization of this goal has generally been hampered, however, by intramolecular and intermolecular vibrational energy redistribution that limits the efficiency of a mode-selective excitation beyond cryogenic temperatures \cite{Uzer1991}. Interest in this field was revitalized recently with the emergence of the field of vibrational polaritonic chemistry \cite{Ebbesen2016a, Feist2018, Flick2018, Ribeiro2018, Thomas2019, Thomas2020, Herrera2020, Hirai2020, Pang2020, Sau2021, Wang2021Perspective,  Garcia-vidal2021, Yang2021}. Changes of reaction rates of molecules inserted into optical infrared cavities have been observed experimentally, demonstrating the influence of the resonant coupling of molecular infrared-active vibrations with the electromagnetic vacuum field of the cavity on the underlying reaction dynamics. Despite intense efforts to understand why the reaction rate changes \cite{Galego2019, Campos-Gonzalez-Angulo2019, Li2020a, Li2021a, Li2021, Schafer2021, Du2021, Mandal2021}, the microscopic mechanisms underlying the experimental observations are still under debate.

Recent studies have hinted at the role of intramolecular vibrational energy redistribution (IVR) in cavity-modified chemical reactions studied \textit{via} first-principles methods \cite{Schafer2021, Head-Marsden2021} or molecular dynamics simulations \cite{Li2020b, Li2021c, Li2021d, Li2021e}. The role of vibrational strong coupling on the chemical reactivity of 1-phenyl-2-trimethylsilylacetylene (PTA) using quantum electrodynamical density functional theory \cite{tokatly2013,Flick2017, Flick2018Nuclear} was able to replicate some of the effects observed experimentally by Thomas \textit{et al.} \cite{Thomas2016}. However, due to the computational complexity, these methods still face certain limitations, such as in the number of simulated trajectories or possible reaction pathways.

Therefore, in this work, we theoretically study cavity-modified unimolecular dissociation reactions with a simple classical model of a molecule. We show that by coupling the infrared-active modes of the molecule to the cavity vacuum field, the rate of bond dissociation in individual molecules can be selectively increased or decreased. In particular, when the cavity is initially empty, the dissociation rate decreases, while when the cavity is initially hotter than the molecule, the cavity can instead accelerate the reaction rate.

\section{Model system} \label{sec:model}

\begin{figure}[!tbhp]
\centering
\includegraphics[width=\linewidth]{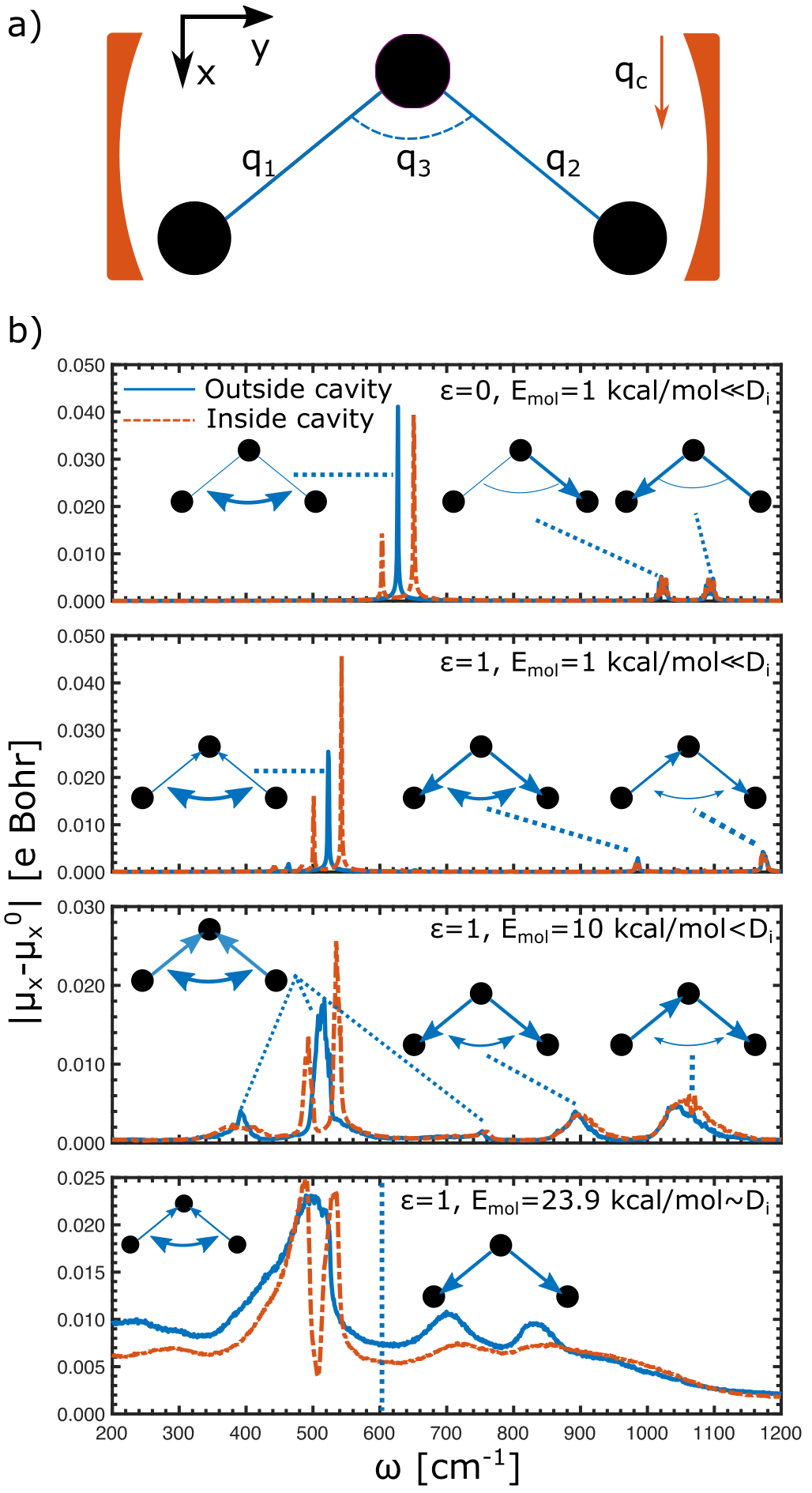}
\caption{Model of a cavity-coupled molecule. \textbf{(a)} Schematic of the model triatomic model in a cavity. The local vibrational degrees of freedom of the molecule are labelled $q_1$ (stretch length between left and center atoms), $q_2$ (stretch length between center and right atoms), and $q_3$ (bending angle between bonds of left-center atoms and center-right atoms), while the cavity mode displacement is $q_\mathrm{c}$. \textbf{(b)} Dipole moment spectrum of the molecule outside and inside the cavity with varying vibrational mode-mode coupling $\epsilon$ and initial energy in the molecule $E_\mathrm{mol}$ and dissociation energy $D_i$. The size and direction of the arrows indicate the relative magnitude and phase of local bond displacement in the normal modes. As the energy increases and more of the anharmonic Morse potentials can be explored, the peaks broaden. We couple the cavity mode to the bending-like modes with the highest intensity and observe Rabi splitting indicative of strong interactions between them.}
\label{fig:bunker}
\end{figure}

We start by discussing the setup of the model system to demonstrate the cavity-modified IVR dynamics. In this work, we use a classical triatomic model with three degrees of freedom and anharmonic potentials. This model was previously used by Karmakar \textit{et al.} \cite{Karmakar2020} to understand the onset of chaos in IVR processes and is itself based on the approach pioneered by D. L. Bunker and others \cite{Bunker1962, Bunker1964, Oxtoby1976} who studied the ozone molecule in a series of computational studies on unimolecular dissociation reactions. Importantly, this minimal model demonstrates chaotic dynamics through an interconnected Arnold web \cite{Froeschle2000, Manikandan2014}, or network of nonlinear resonances, and therefore serves as a fundamental starting point for IVR studies of larger molecules. The Hamiltonian for the model system can be written in terms of the local vibrational modes as
\begin{equation} \label{eq:bunker}
    H_{\mathrm{mol}} = \sum_{i=1}^3\left[\frac{1}{2}G_{ii}^{(0)}p_{i}^2 + V_i(q_i)\right] + \epsilon \sum_{i<j=2}^3  G_{ij}^{(0)} p_i p_j,
\end{equation}
where $i \in \{1, 2, 3\}$ corresponds to the local stretching mode between the left atom and the center atom, the local stretching mode between the center atom and the right atom, and the local bending mode, respectively. The position coordinates $q_i$ for $i=1$ and 2 are in units of distance, while $q_3$ is in units of angle, and $p_i$ are their respective conjugated momentum coordinates. A schematic of the model system and the coordinates is depicted in Fig. \ref{fig:bunker}(a). The coupling element $G_{ij}^{(0)}$ between momenta $p_i$ and $p_j$ is due to the transformation from normal mode coordinates, where we consider collective motions of nuclei, to local mode coordinates, where we focus on the movement of specific parts of the molecule \cite{Cross1933}. To explore the effect of varying vibrational mode-mode momentum couplings, this momentum-momentum interaction is scaled by $\epsilon$, roughly corresponding to \textit{e.g.} varying the mass of the center atom relative to the other two atoms \cite{Bunker1962}. For non-zero $\epsilon$, the local vibrational modes $q_1, q_2$, and $q_3$ do not correspond to normal (or eigen-) modes of the system.  In addition, note that the coordinate-dependent $G_{ij}$ in the original model of Bunker \cite{Bunker1962} are approximated by their equilibrium values $G_{ij}^{(0)}$, and that this model is fixed in the $x$-$y$ plane with dipole activity only in the $x$- and $y$-directions, although we note that the rovibronic couplings have also been shown to impact IVR in certain circumstances \cite{Uzer1985}. For the two stretching modes $i=1, 2$, the vibrational potential is approximated by a Morse potential: $V_i(q_i) = D_i\left\lbrace 1-\exp{\left[-\alpha_i \left(q_i - q_i^0\right)\right]}\right\rbrace^2$, where $D_i$ is the dissociation energy, $\alpha_i$ controls the anharmonicity of the potential well and the number of bound vibrational states in a quantum picture, $q_i^0$ is the equilibrium atom-atom distance, and $\omega_i$ is the frequency of the potential well under the low-displacement or harmonic approximation. For the dissociation energies, we choose $D_1=D_2=D$. For the bending mode, the vibrational potential is assumed to be harmonic: $V_3(q_3) = \omega_3^2 (q_3 - q_3^0)^2/(2G_{33}^{(0)})$. The numerical values for these parameters are defined in Table  \ref{tab:equilibrium} in the Appendix and in Ref. \citenum{Karmakar2020}.

To incorporate the effect of light-matter coupling between the electric field of an infrared cavity and the vibrations of the molecular model, we add the classical field Hamiltonian $H_{\mathrm F}$ \cite{Miller1978}, which becomes the Pauli-Fierz Hamiltonian in its quantized form \cite{Flick2017}:
\begin{equation} \label{eq:LiFieldHamiltonian}
    H_\mathrm{F} = \sum_{k} \left[\frac{1}{2}p_{k}^2 +\frac{1}{2}\omega_k^2 \left(q_{k}-\frac{\lambda_k}{\omega_k}\bm{\mu} \cdot \bm{\xi}\right)^2\right],
\end{equation}
where $\omega_k$, $q_{k}$, $p_{k}$, $\lambda_k$ are the frequency, displacement, momentum, and strength of the cavity mode $k$; $\bm{\mu}$ is the permanent dipole moment of the molecule; and $\bm{\xi}$ is the unit vector of polarization of the electric field. We note that a similar approach has been taken in classical molecular dynamics studies of polaritonic chemistry \cite{Li2020b, Li2021c}. In the following, we will consider interactions with a single photon mode. Further, the permanent dipole moment of the molecule is defined by choosing the $x$-axis to be in-plane and through the center atom along $q_3^0/2$, as in Fig. \ref{fig:bunker}(a). Thus, we parametrize the permanent dipole moment as $\bm{\mu}=A [\cos{(q_3/2)}(q_1+q_2) \hat{x} + \sin{(q_3/2)}(q_1-q_2)\hat{y}]$, where $\hat{x}$ and $\hat{y}$ are the corresponding unit vectors. We set $A=0.4$ e, resulting in the equilibrium permanent dipole moment $\bm{\mu}_0$ in the $x$-direction being 1 e$\cdot$Bohr, on the order of the permanent dipole moment for bent triatomic molecules such as ozone \cite{Adler-Golden1985, Shostak1991}.

For the full light-matter Hamiltonian in the form of  $H=H_{\mathrm{mol}}+H_\mathrm{F}$, the equations of motion can be derived with Hamilton's equations and stably propagated given a set of initial position and momentum coordinates up to several tens of picoseconds using the 8th order Runge-Kutta method \cite{Dormand1981}. To understand the IVR dynamics of the molecule both outside and inside the cavity, we calculate the dipole moment spectrum $|\mu_x-\mu_x^0|(\omega)=|\int \mathrm{d}t (\mu_x(t)-\mu_x^0)\mathrm{exp}(-2\pi\mathrm{i}\omega t)|$. This expression corresponds to the one-sided Fourier transform of the permanent dipole moment $\mu_x$ as the displacements $q_i$ and momenta $p_i$ propagate under $H$ and is directly related to the dipole moment spectrum~\cite{Andrade2009}. Because the IVR dynamics are chaotic \cite{Bunker1964, Oxtoby1976}, all results based on dynamics shown in this study are averaged over tens of thousands of initial states that are generated as described in the Appendix.

We first describe the molecule outside of the cavity. In the simplest case, shown in the top plot of \fig{fig:bunker}(b), there is no mode-mode momentum coupling, \textit{i.e.} $\epsilon=0$ in Eq.~\ref{eq:bunker}, and the initial energy of 1 kcal/mol is low compared to the dissociation energy $D=24 $kcal/mol of the local stretching modes and of the same order of magnitude as their harmonic frequencies. Therefore, the Hamiltonian effectively corresponds to three decoupled local vibrations. Notably, the local bending peak at 632 cm$^{-1}$ is sharper and brighter, while the local stretches at 1040 and 1112 cm$^{-1}$ are broader and darker. The difference in peak widths is a result of the anharmonic local stretching potentials. In the second plot of \fig{fig:bunker}(b), we turn on the vibrational mode-mode momentum coupling by setting $\epsilon = 1$. As expected, we find that the local vibrational modes are no longer the normal modes of the system---the normal modes are now delocalized over the three local modes, shifting the three center frequencies $(\omega_1, \omega_2, \omega_3)$ of the normal bending, normal symmetric stretching, and normal anti-symmetric stretching modes to $(1197, 1003, 528)$ cm$^{-1}$ from $(1112, 1040, 632)$ cm$^{-1}$ in the case of $\epsilon=0$.
As the energy of the initial state is increased to 10 kcal/mol in the third plot of Fig. \ref{fig:bunker}(b), such that $q_1$ and $q_2$ can explore more of the anharmonic Morse potential, we observe the emergence of additional and broadened peaks corresponding to anharmonic resonances. Nonetheless, we can still identify peaks with frequencies below or above $\sim$800 cm$^{-1}$ with the normal bending mode or normal stretching modes, respectively. As the energy of the initial states is increased to 23.9 kcal/mol in Fig. \ref{fig:bunker}(b), just below the dissociation energy $D=24$ kcal/mol, the frequency spectra of the local stretching modes $q_1$ and $q_2$ have largely delocalized across the entire frequency spectrum with a broad peak near $\sim$500 cm$^{-1}$ still corresponding to normal bending mode-like character.

For the spectra including the optical cavity mode in Fig.~\ref{fig:bunker}, we couple the molecule to a single cavity mode $c$ that we assume is $x$-polarized. The photon mode is then initialized with vanishingly small energy. We resonantly tune the cavity mode to the frequency with the largest intensity in the $|\mu_x-\mu_x^0|(\omega)$ spectrum that corresponds to a bending mode and choose the cavity strength $\lambda_c=0.05$ a.u. leading to an estimated Rabi splitting $\hbar\Omega_\mathrm{R}=45$ cm$^{-1}$ or 1.35 THz. This Rabi splitting is the border between the strong and ultra-strong coupling regimes. (We describe how the Rabi splitting can be estimated in the low-energy harmonic case in the Appendix.) Note that this cavity strength implies a much smaller mode volume $V$ than experimental values to compensate for the lack of $N$ molecules in this single-molecule model, since $\Omega_\mathrm{R}\propto \sqrt{N/V}$. The peak splitting is clear evidence of interactions between the cavity mode and vibrations, suggesting energy can be transferred between the cavity mode and vibrational modes, and the magnitude of the splitting is similar regardless of $\epsilon$ and $E_\mathrm{mol}$.

\section{Results}

\subsection{Dissociation Dynamics}

\begin{figure}[!tbhp]
\centering
\includegraphics[width=\linewidth]{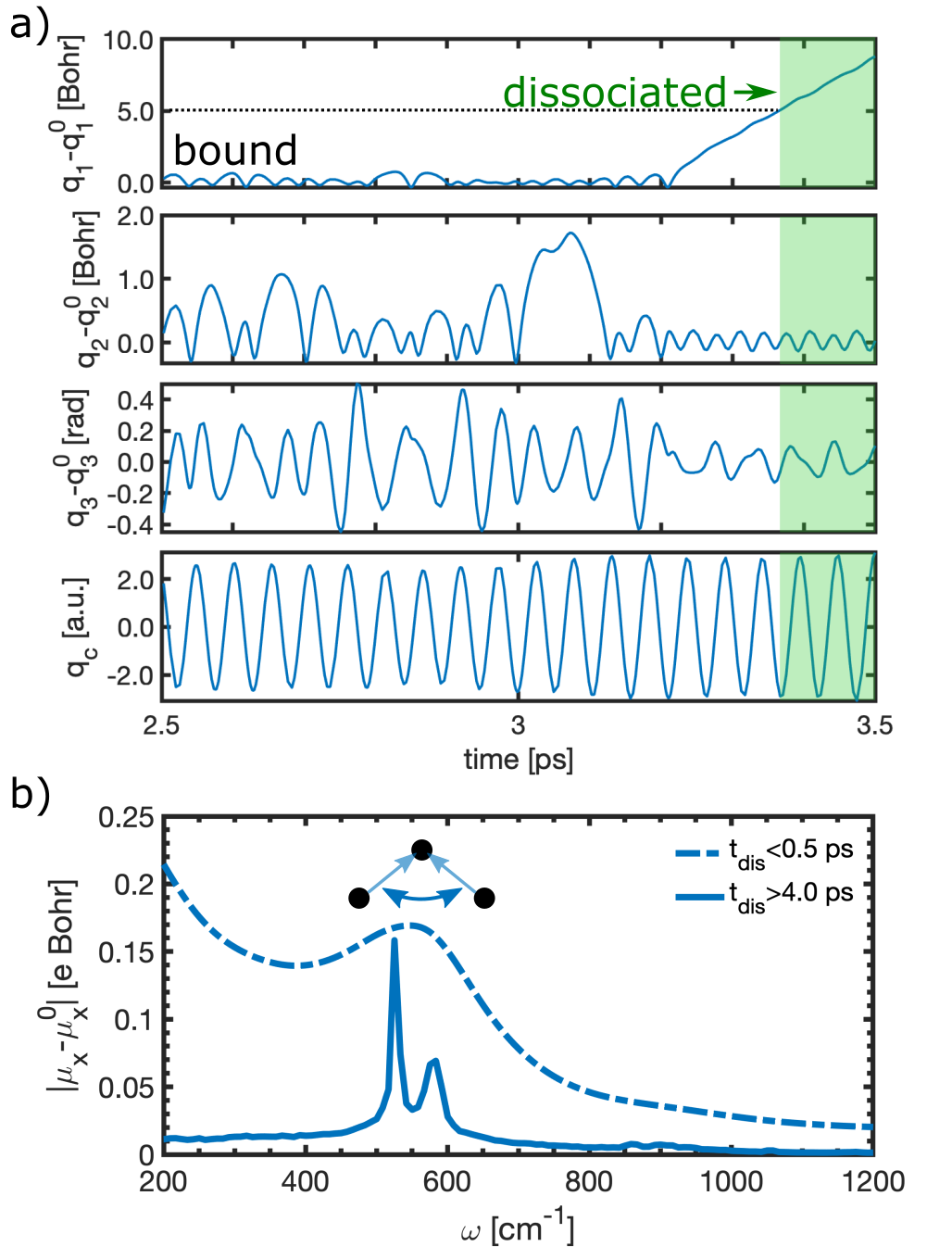}
\caption{IVR-mediated dissociation. \textbf{(a)} Single trajectory of a molecule initialized with energy of 34 kcal/mol that is greater than the dissociation energy $D_i=24$ kcal/mol. Local mode displacements $q_i-q_i^0$ are in atomic units. The molecule dissociates at $\sim$3.4 ps, marked by the green region, when $q_1-q_1^0$ exceeds dissociation length $r_{\mathrm{dis}}=5$ Bohr. \textbf{(b)} Dipole moment spectrum of short-lived (dissociation time smaller than 0.5 ps) and long-lived (dissociation time greater than 4 ps) trajectories of the molecule outside the cavity. Long-lived trajectories exhibit sharp resonances, suggesting that strongly coupling the cavity mode to these particular modes can influence dissociation dynamics.}
\label{fig:dissociation}
\end{figure}

Next, we study the dissociation dynamics of the molecular system. Due to the classicality of the model, the molecule can dissociate when it is initialized with energy exceeding the dissociation energy $D_i=24$ kcal/mol. We define successful dissociation when either $q_1$ or $q_2$ exceeds a bond length of 5 Bohr from the equilibrium value $q_1^0$ or $q_2^0$. Setting $E_\mathrm{mol}=34$ kcal/mol, we show an example trajectory that leads to dissociation of the molecule inside a cavity in Fig. \ref{fig:dissociation}(a), where we plot the deviation of the molecular coordinates $q_i$ from their equilibrium positions $q_i^0$. The trajectory exhibits anharmonic oscillations of the internal coordinates of the molecule resulting finally in the divergence of the bond length $q_1$ and, therefore, dissociation of the molecule.  

We generate a statistical understanding of the dissociation dynamics by studying tens of thousands of trajectories with identical energies. In \fig{fig:dissociation}(b), we plot the dipole moment spectrum $|\mu_x-\mu_x^0|(\omega)$ for molecules initialized with energy 34 kcal/mol $> D=24$ kcal/mol, sufficient to dissociate the bond. Binning trajectories by their lifetimes, we compare the dipole moment spectra of short-lived trajectories ($t_\mathrm{dis}<0.5$ ps) in blue and long-lived trajectories ($t_\mathrm{dis}>4$ ps) in orange. For the short-lived trajectories we observe uniform spectral densities that hint at statistical decay dynamics in which the many anharmonic resonances of the local vibrational modes are explored quickly and at random. In contrast, the Fourier spectra of the long-lived trajectories feature a number of peaks; especially prominent ones are at $\sim$520 cm$^{-1}$ and $\sim$580 cm$^{-1}$. These peaks represent modes robust against dissociation, where the molecular energy is largely stored in the local bending motion and not local stretches prone to dissociation. Therefore, we suggest to alter the dynamics of IVR-mediated dissociation by coupling the cavity to these resonances.

\subsection{Cavity-modified dissociation} \label{sec:cavity}

\begin{figure}[!tbhp]
\centering
\includegraphics[width=1.0\linewidth]{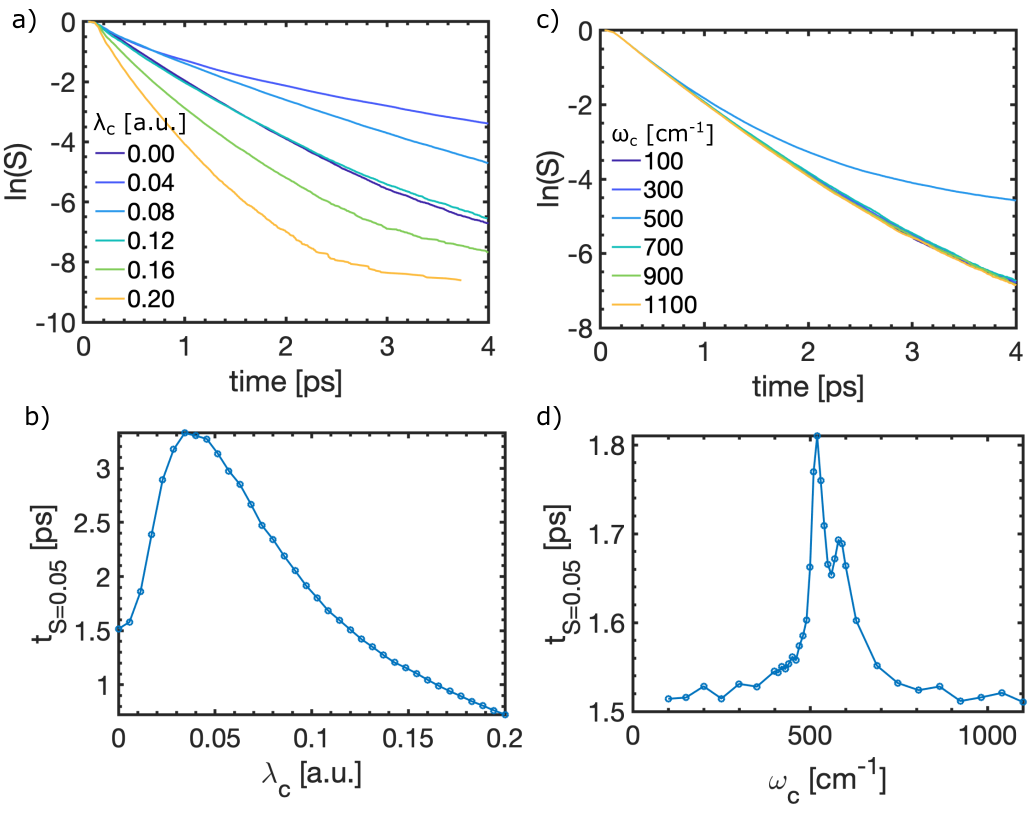}
\caption{Cavity-modified unimolecular dissociation rate with survival probability as a function of time (top) and time required for 95\% of molecules with initial energy of 34 kcal/mol (bottom) for an initially empty cavity. In \textbf{(a)-(b)}, we vary the cavity strength $\lambda_{\mathrm c}$ for $\omega_{\mathrm c}=520$ cm$^{-1}$ and $x$-polarized cavity mode, resulting in the dissociation rate slowing down the most around $\lambda_\mathrm{c}=0.04$ a.u. before decreasing with further increasing $\lambda_\mathrm{c}$. In \textbf{(c)-(d)}, the cavity frequency $\omega_{\mathrm c}$ is varied for $\lambda_{\mathrm c}=0.01$ a.u. and $x$-polarized cavity mode; the dissociation rate slows down between $\omega_c\sim$520 and $\sim$590 cm$^{-1}$, the same frequencies of the peaks in $|\mu_x-\mu_x^0|$ in Fig. \ref{fig:dissociation}(b).}
\label{fig:reactionrate}
\end{figure}

We explore whether coupling the molecule to the cavity can modify reaction dynamics. In our model, we initialize the molecule with $E_\mathrm{mol}=34$ kcal/mol, set $\epsilon=1$, and couple it to an empty single cavity mode with frequency $\omega_{\mathrm c}$ and with cavity strength $\lambda_{\mathrm c}$ oriented along the direction $\theta$ from the $x$-axis.

First, we calculate the change in reaction rate for varying cavity strength $\lambda_\mathrm{c}$. To set $\omega_\mathrm{c}$ and $\theta$, we note that $|\mu_x-\mu_x^0|(\omega)$ is largest around 520 cm$^{-1}$ in Fig. \ref{fig:dissociation}(b). Therefore, we presume the strength of light-matter coupling for a given $\lambda_{\mathrm c}$ can be maximized by polarizing the cavity field along the $x$-direction and setting $\omega_{\mathrm c}=520$ cm$^{-1}$. We vary the cavity strength $\lambda_{\mathrm c}$ and plot the (natural logarithm of) the survival probability $S(t)$, or the proportion of initial states that have not dissociated as a function of time in Fig.\,\ref{fig:reactionrate}(a), as well as the time required for 95\% of the molecules to dissociate in Fig.\,\ref{fig:reactionrate}(b). Importantly, from $\lambda_\mathrm{c}=0$ to around 0.04 a.u., the dissociate slows monotonically with increasing $\lambda_\mathrm{c}$. As $\lambda_\mathrm{c}$ increases further above 0.04 a.u. into the ultra-strong coupling regime where the dipole self-energy causes the total system energy to increase substantially above the 34 kcal/mol initial energy of the bare molecule, the reaction accelerates to become faster than even outside the cavity.

We further explore the cavity-coupled triatomic model by varying cavity frequency $\omega_{\mathrm c}$ from 100 cm$^{-1}$ to 1100 cm$^{-1}$ with constant cavity strength $\lambda_{\mathrm c}=0.01$ a.u. From Fig. \ref{fig:reactionrate}(c), we see that for much of this frequency range, the survival probability $S(t)$ curves are nearly identical to the case of $\lambda_{\mathrm c}=0$ in Fig. \ref{fig:reactionrate}(a). Around 500 cm$^{-1}$, however, the reaction dynamics are strongly changed. To study them with a finer frequency resolution in this range, we plot the time required for 95\% dissociation in Fig. \ref{fig:reactionrate}(d), where we observe peaks at $\sim$520 cm$^{-1}$ and $\sim$580 cm$^{-1}$, the same frequencies of the peaks $|\mu_x - \mu_x^0|(\omega)$ in Fig. \ref{fig:dissociation}(b) and therefore serving as direct evidence of a resonant effect. We show results for the case of a $y$-polarized cavity in the Appendix and for molecules with weak vibrational mode-mode coupling $\epsilon$ in the Appendix, which exhibits reaction acceleration inside a cavity with $\lambda_\mathrm{c}=0.01$ a.u.

To develop a qualitative understanding of how the cavity changes the reaction rate, we study the change in the form of the decay curves as the reaction time $t_{S=0.05}$ increases for both varying $\lambda_\mathrm{c}$ and $\omega_\mathrm{c}$. For $\lambda_{\mathrm c} = 0$ corresponding to complete decoupling between the molecule and the cavity, the survival probability $S(t)$ can be described with a mono-exponential decay process \cite{Karmakar2020}, corresponding to statistical distribution of vibrational energy where the space of nonlinear frequency resonances between vibrational modes is widely explored before the molecule dissociates. The reaction slows because a secondary, slower decay channel becomes more dominant, veering the survival probability $S(t)$ away from mono-exponential statistical decay. This slower decay channel has been shown to correspond to IVR-limited dissociation, where the molecule vibrates along particular modes before dissociating \cite{Karmakar2020}, as shown by the dipole moment spectrum of short- vs long-lived trajectories in Fig. \ref{fig:dissociation}(b). 

\begin{figure}[!tbhp]
\centering
\includegraphics[width=1.0\linewidth]{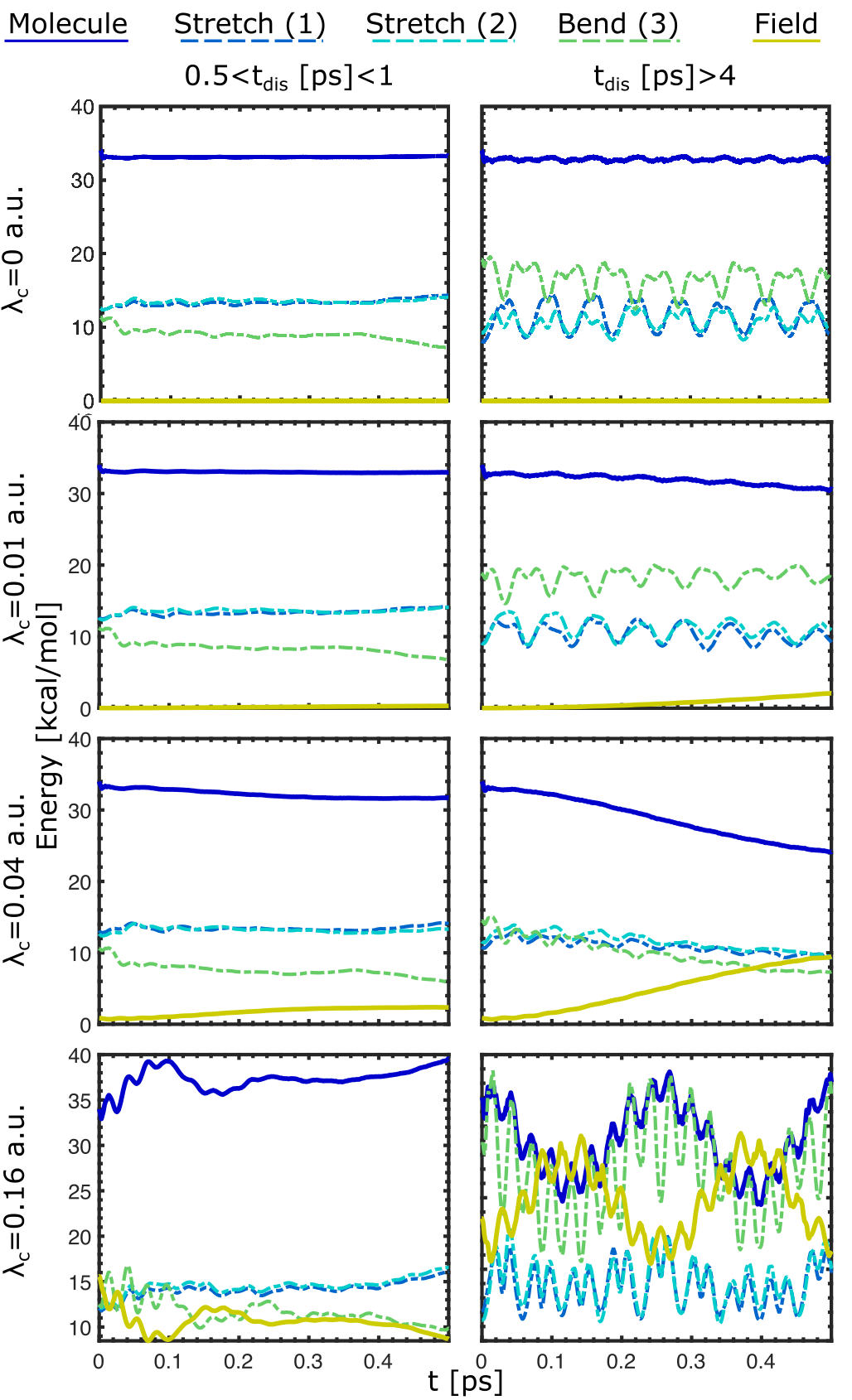}
\caption{Time-dependent energy distribution of the first 0.5 ps in the molecule-cavity system for short-lived ($0.5<t_\mathrm{dis}<1$ ps) and long-lived ($t_\mathrm{dis}>4$ ps) trajectories where $\lambda_\mathrm{c} \in \{0, 0.01, 0.04, 0.16\}$ a.u. The resonantly tuned cavity mode couples efficiently to molecules with highly excited local bending modes and functions as a reservoir of molecular vibrational energy. The reaction rate increases for increasing coupling strength $\lambda_\mathrm{c}$ above 0.04 a.u. as the vibrational energy is quickly returned to the molecule and the dipole self-energy induced by the cavity increases the energy of the molecule.}
\label{fig:reservoir}
\end{figure}

We demonstrate that the cavity increases the number of trajectories that decay \textit{via} the IVR-limited decay channel by serving as a sink for vibrational energy, lowering the probability that one of the stretch bonds acquires enough energy to break. In Fig. \ref{fig:reservoir}, for different cavity strength $\lambda_\mathrm{c} \in \{0, 0.01, 0.04, 0.16\}$ a.u., we plot the time-dependent energy distribution up to the first 0.5 ps in the molecule-cavity system for shorter-lived ($0.5<t_\mathrm{dis}<1$ ps) and longer-lived ($t_\mathrm{dis}>4$ ps) trajectories inside and outside the cavity averaged over $\sim$200,000 initial states. The distribution comprises energy in the entire molecule as $H_\mathrm{mol}$, the field $H_\mathrm{F}$ (including matter-photon coupling), and each of the local vibrational modes $i$ as $\frac{1}{2}G_{ii}^{(0)}p_{i}^2 + V_i(q_i)$. Note that for clarity, we do not show the vibrational mode-mode coupling, which is generally unaltered by the presence of the cavity. 

At $\lambda_\mathrm{c}=0$ corresponding to the molecule being outside the cavity, for the shorter-lived trajectories, energy is initially distributed approximately evenly between the local bending and local stretching modes and is gradually redistributed from the local bending to the local stretching modes. In sharp contrast, for the longer-lived trajectories the initial states are more highly excited in the local bending mode. Importantly, the energy oscillations indicate localization in frequency space within the sharp resonances of the $x$-polarized dipole moment spectra around 500-600 cm$^{-1}$.

Inside the cavity with $\lambda_\mathrm{c}=0.01$ a.u., for both shorter- and longer-lived trajectories, we see that the cavity mode strongly couples to the local bending mode, as the field energy increases while the molecular energy decreases. Notably, the local bending mode is more excited for the longer-lived trajectories. The strongly coupled cavity mode is then able to absorb more energy from both the local bending mode and local stretching modes, in turn extending the average lifetime of the molecule.

We also explain why the reaction rate initially increases with increasing cavity strength $\lambda_\mathrm{c}$ until $\lambda_\mathrm{c}\sim 0.04$ a.u. and then decreases with further increasing $\lambda_\mathrm{c}$. Up to $\lambda_\mathrm{c}=0.04$ a.u., the reaction slows with increasing $\lambda_\mathrm{c}$ as the cavity absorbs vibrational energy. Beyond $\lambda_\mathrm{c}=0.04$ a.u., such as at $\lambda_\mathrm{c}=0.16$ a.u., the reaction rate increases as the cavity initially gives energy to the molecule for short-lived trajectories and quickly returns vibrational energy to the molecule for longer-lived trajectories. The non-zero initial energy of the field at $\lambda_\mathrm{c}=0.16$ a.u. is due solely to the matter-photon coupling, and in this case the dipole self-energy, as the initial cavity state is vanishingly empty. Overall, we find that the cavity is best able to slow down the reaction by quickly absorbing energy from the molecule without returning it too quickly, highlighting an important subtlety in cavity-modified IVR.

\begin{figure}[!tbhp]
\centering
\includegraphics[width=1.0\linewidth]{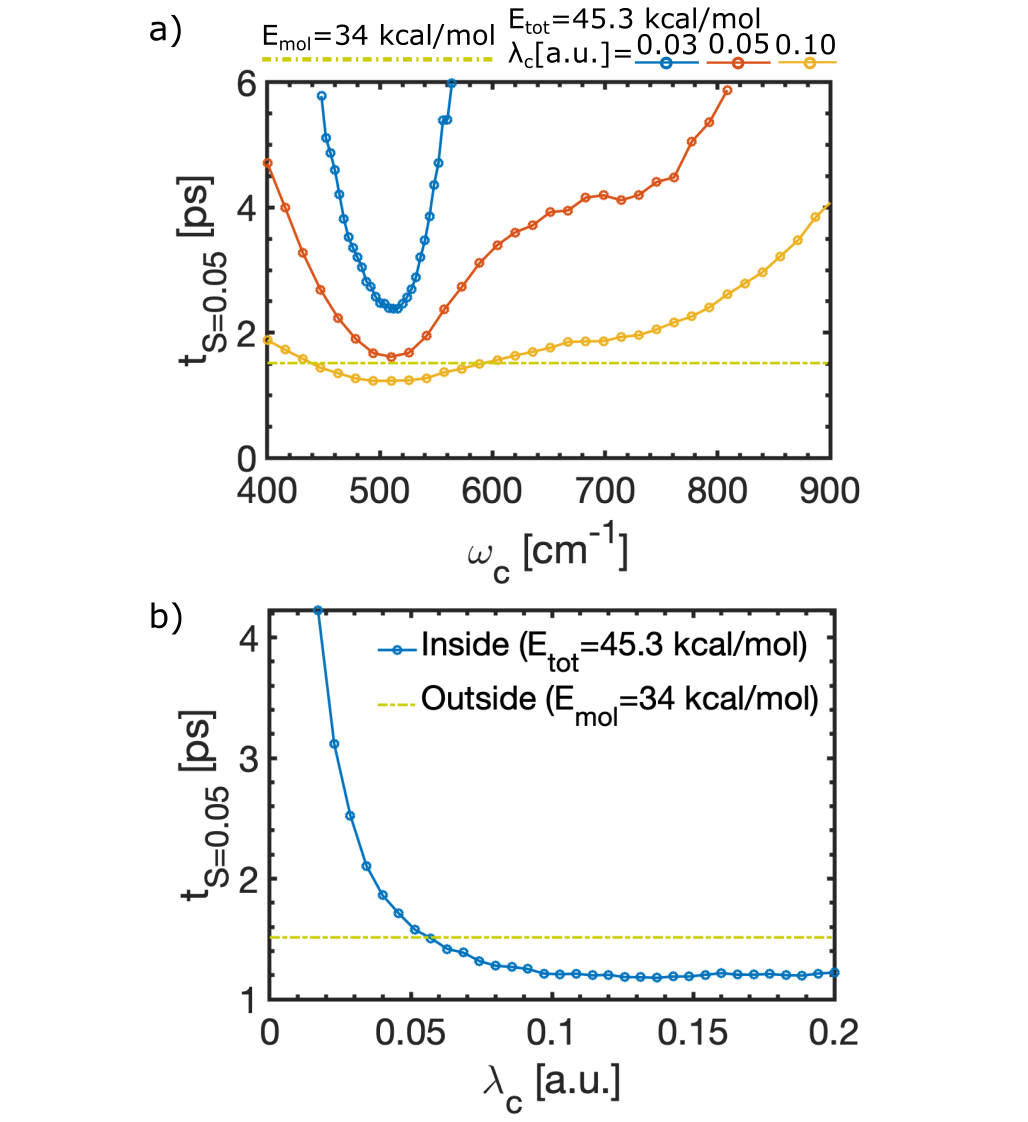}
\caption{Dissociation time $t_{S=0.05}$ inside a hot cavity with varying \textbf{(a)} cavity frequency $\omega_\mathrm{c}$ and \textbf{(b)} cavity strength $\lambda_\mathrm{c}$ at the resonant cavity frequency $\omega_\mathrm{c}=520$ cm$^{-1}$. The total molecule-cavity energy is $34\cdot4/3\sim 45$ kcal/mol, where the factor of $4/3$ accounts for the additional degree of freedom by the cavity mode. As a guide to the eye, we also plot $t_{S=0.05}$ for the bare molecule with 34 kcal/mol (segmented yellow). The resonance is broader for the hot cavity than for the empty cavity in Fig. \ref{fig:reactionrate}, and the reaction rate for a given $\lambda_\mathrm{c}$ is fastest at the resonance. The reaction rate decreases toward a plateau that is faster than the rate outside the cavity beyond $\lambda_\mathrm{c}=0.05$ a.u. At off-resonant cavity frequencies and when cavity-molecule interactions are weak, energy localized in the cavity cannot be efficiently transported to the molecule.}
\label{fig:temperature}
\end{figure}

The cavity mode, initially vanishingly empty in Fig. \ref{fig:reactionrate} and Fig. \ref{fig:reservoir}, appears to slow down the dissociation rate by absorbing energy from the vibrationally excited molecule. In the case of chemical reactions at thermal equilibrium, however, the cavity mode is likely to be thermally occupied. Therefore, in Fig. \ref{fig:temperature}, we present the effect of a ``hot" cavity on dissociation dynamics, where the total molecule-cavity energy is $34\cdot4/3\sim 45$ kcal/mol. The factor of $4/3$ takes into account the additional degree of freedom by the cavity mode such that the average energy per degree of freedom is the same for the molecule inside and outside the cavity. 

Below the cavity strength $\lambda_\mathrm{c} < 0.05$ a.u., the reaction rate is slower inside the cavity for all cavity frequencies, reaching its fastest rate close to the resonant frequency found in Fig. \ref{fig:reactionrate}(d). As the cavity strength $\lambda_\mathrm{c}$ increases beyond 0.05 a.u. toward the ultra-strong coupling regime, the reaction can be even faster inside the cavity than outside. These results can be understood as follows: off-resonant and weak cavity-molecule interactions can prevent energy localized in the cavity from transporting to the molecule. 

\begin{figure*}[!tbhp]
\centering
\includegraphics[width=1.0\linewidth]{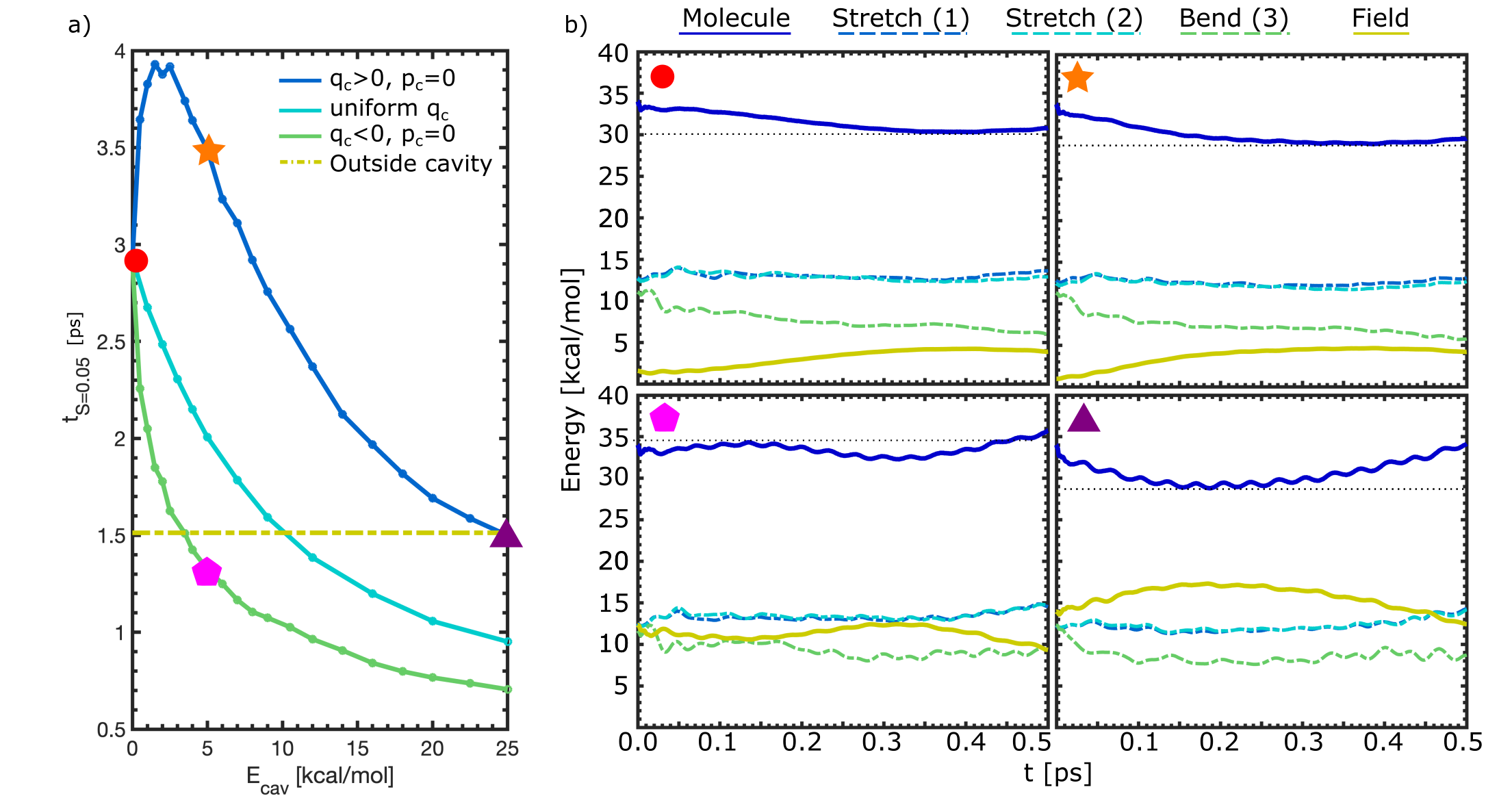}
\caption{Reaction rate dependence on cavity state. \textbf{(a)} Time $t_{S=0.05}$ for 95\% of molecules to dissociate for varying cavity energy $E_\mathrm{cav}$, where the cavity is initialized with $q_\mathrm{c}>0$ and $p_\mathrm{c}=0$ (dark blue), uniformly sampled $q_\mathrm{c}$ and $p_\mathrm{c}$ (``uniform $q_\mathrm{c}$" in light blue), and $q_\mathrm{c}<0$ and $p_\mathrm{c}=0$ (green). In \textbf{(b)}, we plot the time-dependent energy distribution for the colored shapes in (a). The local maximum in $t_{S=0.05}$ for $q_\mathrm{c}>0$ is a result of competition between absorbing more energy from the molecule and the cavity returning it quickly. $q_\mathrm{c}<0$ (pink) results in faster reactions than $q_\mathrm{c}>0$ (orange) at the same $E_\mathrm{cav}$ because in the former, energy is initially transferred from the cavity to the molecule. The black, dotted line marks the minimum (maximum) energy of the molecule after the first oscillation of energy transfer to (from) the cavity.}
\label{fig:cavitystate}
\end{figure*}

To control for the effects of initial states where the molecule energy is too low to dissociate, in Fig. \ref{fig:cavitystate} we fix the energy of the molecule to be the same as it is outside the cavity and study how increasing cavity energy affects the reaction rate. We find that hotter cavities generally increase the reaction rate, although the exact phase of the cavity state relative to the molecule's can drastically affect the reaction rate.

To explore whether certain initial cavity states, which could potentially be preferentially generated by non-equilibrium effects such as coherent pumping, can affect dissociation dynamics differently, we explore the following different regimes in Fig.~\ref{fig:cavitystate}: $q_\mathrm{c}>0$ and $p_\mathrm{c}=0$ (dark blue), $q_\mathrm{c}<0$ and $p_\mathrm{c}=0$ (green), and uniformly sampled $q_\mathrm{c}$ and $p_\mathrm{c}$ (light blue). Further details on the initialization of the cavity states are presented in the Appendix. Different initial signs of $q_\mathrm{c}$ could result in qualitatively different effects because the angle $\theta$ of the molecule orientation from the  breaks a symmetry of the system, such that for positive (negative) $q_\mathrm{c}$, the light-matter interaction is negative (positive). Then, for each of these states, in Fig. \ref{fig:cavitystate}(a) we calculate the reaction time $t_{S=0.05}$ of molecules with fixed initial energy 34 kcal/mol for varying initial cavity energy $E_\mathrm{cav}=p_\mathrm{c}^2/2+\omega_\mathrm{c} q_\mathrm{c}^2/2$. 

For ``uniform $q_\mathrm{c}$" and $q_\mathrm{c}<0$, as $E_\mathrm{cav}$ increases, we find that the reaction rate decreases and can become even faster than the reaction outside the cavity (dotted yellow). Notably, the reaction rate of the ``uniform $q_\mathrm{c}$"  case with $E_\mathrm{cav}\sim$10 kcal/mol equals the reaction rate outside the cavity, where both cases have approximately the same amount of energy per degree of freedom. In contrast, for $q_\mathrm{c}>0$, the reaction rate slows down further from $E_\mathrm{cav}=0$ to $\approx 3$ kcal/mol and decreases the rate outside the cavity with further increasing $E_\mathrm{cav}$ until $E_\mathrm{cav}$ is as high as 25 kcal/mol.

To explain these results, in Fig. \ref{fig:cavitystate}(b), we plot the time-dependent energy distribution for the conditions marked by the colored circles in (a). For $q_\mathrm{c}>0$, increasing $q_\mathrm{c}$ with cavity energies $E_\mathrm{cav}$ from nearly 0 (red) to 5 (orange) to 25 (purple) kcal/mol results in two competing effects: the maximum amount of energy the cavity can absorb from the molecule increases while the rate at which the cavity returns the energy to the molecule also increases. The competition between these two processes results in the local maximum in $t_{S=0.01}$ for $q_\mathrm{c}>0$. $q_\mathrm{c}<0$ (pink) results in faster reactions than $q_\mathrm{c}>0$ (orange) at the same energy because in the former, energy is initially transferred from the cavity to the molecule. The initial direction of energy transfer differs between positive and negative $q_\mathrm{c}$ because the sign of $q_\mathrm{c}$ determines the overall sign of the light-matter interaction term $\omega_\mathrm{c}\mu\lambda_\mathrm{c}q_\mathrm{c}$, which then affects the initial sign and magnitude of the time derivatives of the molecule-cavity coordinates. Overall, these results suggest that cavities even ``hotter" than the ``hot" molecules can appreciably affect dissociation dynamics if the state of the cavity can be controlled, potentially with coherent pumping.

\section{Conclusions and outlook}

In summary, we study the dissociation dynamics of a classical model of a triatomic molecule in an optical cavity based on a Bunker model, where the potential of the two local stretches are anharmonic Morse potentials and the local bending mode is harmonic. We extend this model by resonantly coupling the vibrational states with large optical dipole moments to a single electromagnetic mode of an infrared cavity. Importantly, we observe that the effect of the cavity is strongest when it is tuned into resonance with the frequencies of dissociation-resistant vibrational modes. In addition, we find that the reaction rate monotonically decreases until the molecule-cavity system enters the ultra-strong coupling regime. With further increasing light-matter interaction, the reaction rate increases and can become even faster in the cavity than outside of it. We find that the reaction is maximally slowed down when the cavity is able to absorb energy from the molecule without returning it too quickly to the molecule, lowering the probability that one of the stretch bonds acquires enough energy to break, and the reaction speeds up in the ultra-strong regime due to the additional energy from the dipole self-energy term.

When the energy of the molecule-cavity system is increased such that the energy per degree of freedom is equivalent to that of the bare molecule, the reaction rate is generally slowed down as energy initialized in the cavity is unable to be transferred to the molecule without resonant and strong cavity-molecule interaction. To control for this localization effect, we also fix the molecule energy and study dissociation as a function of the cavity energy. We find that cavity-modified reaction rates are generally only observable when the cavity is ``cooler" compared to the molecule. Nevertheless, even a ``hot" cavity can slow down reaction dynamics. As an example, we show that particular cavity states are more prone to initially absorbing energy from the molecule, while others with the same initial energy are more prone to transferring energy to the molecule, both of which are a result of the initial sign of the light-matter interaction term. These results highlight the importance of dynamical effects in vibrational polariton chemistry.

This study lays the foundation for further theoretical and experimental work toward understanding the intriguing experimental results of vibrational polariton chemistry. Regarding theoretical work, we expect molecular dynamics studies of many molecules inside an optical cavity that also include molecular translational and rotational degrees of freedom and intramolecular interactions \cite{Luk2017, Groenhof2019, Li2020b, Li2021c, Li2021d, Li2021e} with a particular focus on IVR dynamics to be fruitful. In addition, in cases when the dissociation energy becomes comparable to a quantum of vibrational energy, quantum effects can play an important role in the dissociation dynamics \cite{Thomas2016}. In such cases a quantum description, e.g. \textit{via} first-principles method such as quantum electrodynamical density functional theory \cite{tokatly2013, ruggenthaler2014, Flick2017, Flick2018Nuclear, Wang2021a, Schafer2021} is necessary. While for computational feasibility and ease of interpretation, we consider only a triatomic anharmonic model that serves as a minimal viable model for chaotic dynamics, molecules studied in experiments are considerably more complex. They can contain $\sim$100 vibrational degrees of freedom. These many degrees of freedom may exhibit vibrational mode-mode couplings that span several orders of magnitude. Here, we also do not consider cavity losses, as they are estimated to be significantly slower than IVR dynamics at $\sim$5 ps \cite{Li2021d}. Future studies should therefore determine whether the observed cavity-modified chemical reactivity in this study is robust to these factors. Finally, experimental studies that consider reactions limited by IVR that therefore do not abide by the standard assumptions of transition rate theory and where the initial states of the intramolecular vibrational and cavity degrees of freedom can be controlled will help to further clarify the role of IVR in vibrational polaritonic chemistry.

\section*{Associated content}
Code to reproduce the calculations in this paper are available at \href{https://github.com/drekwang/cavityIVR}{https://github.com/drekwang/cavityIVR}.

\section*{Acknowledgements}
S.F.Y. and J.F. contributed equally to this manuscript. D.S.W. acknowledges valuable discussions with Arkajit Mandal, Srihari Keshavamurthy, Sourav Karmakar, Subhadip Mondal, Joel Yuen-Zhou, Matthew Du, Jorge Campos-Gonzalez-Angulo, Eric J. Heller, Dominik Sidler, and Vamsi V. Varanasi. D.S.W. is an NSF Graduate Research Fellow. S.F.Y. would like to acknowledge funding by DOE, AFOSR, and NSF. Calculations were performed using the computational facilities of the Flatiron Institute. The Flatiron Institute is a division of the Simons Foundation. 

\appendix

\section{Model parameters} \label{app:parameters}

\begin{table}[!htp]
    \centering
    \begin{tabular}{||c|c|c|c||}
        \hline
         $q_1^0$ &  $q_2^0$ & $q_3^0$ & \\
         \hline
         2.426 & 2.426 & 2.039 & \\
         \hline
        \hline
         $\alpha_1$ &  $\alpha_2$ &  $D_1$ & $D_2$ \\
         \hline
         2.212 & 2.069 & 24 & 24 \\
         \hline 
         \hline 
         $G_{11}^{(0)}=G_{22}^{(0)}$ & $G_{33}^{(0)}$ & $G_{12}^{(0)}$ & $G_{13}^{(0)}=G_{23}^{(0)}$  \\
         \hline
         $6.85\times 10^{-5}$ & $2.88 \times 10^{-5}$ & $-1.54\times 10^{-5}$ & $1.26 \times 10^{-5}$ \\
         \hline 
    \end{tabular}
    \caption{Parameters for $H_\mathrm{mol}$ (Eq.~\ref{eq:bunker}) in atomic units, except $D_i$ in kcal/mol, adapted from Ref. \citenum{Karmakar2020}. These values result in local mode frequencies of $(\omega_1, \omega_2, \omega_3) = (1112, 1040, 632)$ cm$^{-1}$ for low vibrational excitations corresponding to the harmonic approximation.}
    \label{tab:equilibrium}
\end{table}

\FloatBarrier

\section{Initial states} \label{app:initialstates}

We briefly review the method of generating initial molecular states, as originally described in Ref.~\citenum{Karmakar2020}, and discuss how it is generalized to initialize the cavity. 

For a given molecule energy $E_\mathrm{mol}$, we compute the minimum $Q_\mathrm{min}$ and maximum $Q_\mathrm{max}$ possible values of $m\in \{q_1, q_2, p_1, p_2, p_3\}$ corresponding to the cases when $E_\mathrm{mol}$ is contained entirely within the potential or kinetic energy associated with $Q$. Then, we generate all possible initial states from uniformly spaced arrays with $N$ values between $m_\mathrm{min}$ and $m_\mathrm{max}$. For each of these states with energy less than $E_\mathrm{mol}$, we choose the positive root of $q_3-q_3^0$ that results in a total energy of $E_\mathrm{mol}$. To obtain converged results in this study, we choose $N$ between 15 and 23, where larger values are necessary for survival probabilities and smaller values are sufficient for Fourier spectra and time-dependent energy curves.

To initialize the molecule-cavity system with a given $E_\mathrm{total}$ dictated by $H_\mathrm{mol}+H_\mathrm{F}$, as in Fig. \ref{fig:temperature}, we adapt the procedure from the case of the bare molecule described above, except that we generate all possible initial states from uniformly spaced arrays of $q_1, q_2, q_3, p_1, p_2, p_3, q_\mathrm{c}$ and select the value of $p_\mathrm{c}$ that gives the desired total energy.

To initialize the cavity with a given $E_\mathrm{cav}=p_\mathrm{c}^2/2+\omega_\mathrm{c}^2 q_\mathrm{c}^2 / 2$, as in Fig. \ref{fig:cavitystate}, we first compute the minimum $q_\mathrm{min, c}$ and maximum $q_\mathrm{max,c}$ possible values of $q_c$ corresponding to $E_\mathrm{cav}=\omega_\mathrm{c}^2 q_\mathrm{c}^2 / 2$. For the study of $q_\mathrm{c}>0$ ($q_\mathrm{c}<0$), for each molecule initial state, there is only one initial cavity state with $q_\mathrm{c}=q_\mathrm{max,c}$ ($q_\mathrm{c}=q_\mathrm{min,c}=-q_\mathrm{max,c}$) and $p_\mathrm{c}=0$. For the ``uniform $q_\mathrm{c}$" studies, for each molecule initial state, we generate $N_\mathrm{c}$ cavity states, where $q_\mathrm{c}$ selected from a uniformly spaced array with $N_\mathrm{c}$ points between $q_\mathrm{min, c}$ and $q_\mathrm{max, c}$ and $p_\mathrm{c}$ is chosen such that $E_\mathrm{cav}=p_\mathrm{c}^2/2+\omega_\mathrm{c}^2 q_\mathrm{c}^2 / 2$. Note that when the cavity is initialized as vanishingly empty in Fig. \ref{fig:bunker}, \ref{fig:dissociation}, \ref{fig:cavitystate}, \ref{fig:reservoir}, \ref{fig:rabi}, and \ref{fig:reactionrate_y}, we set $E_\mathrm{cav}$ at a small but non-zero value on the order of 0.1 kcal/mol to ensure numerical stability.

In the present study we generate ensembles of trajectories picked from constant energy hypersurfaces \cite{Karmakar2020} to elucidate the fundamental physics. Future studies could consider differently chosen ensembles of initial states relevant to particular experimental conditions. For instance, to simulate conditions of vibrational excitation at room-temperature, energies and momenta of the molecule and cavity can be selected from a Boltzmann distribution. Alternatively, to simulate population \textit{via} infrared laser pumping, only optically accessible initial states can be chosen.

\section{Estimating the Rabi splitting} \label{app:Rabi}
We demonstrate how to estimate the Rabi splitting induced by the cavity mode in the dipolar moment spectrum of the vibrational modes in a manner similar to Refs. \citenum{Shalabney2015, Galego2019, LiHuo2021}. As an example, we assume that the single cavity mode $c$ is aligned along the in-plane axis through the center of the tri-atomic model to $q_3/2$, such that the only component of the permanent dipole moment $\mu$ relevant to light-matter coupling is given by $A\cos (q_3/2)(q_1+q_2)$, where the coefficient $A$ is fitted with experimental or computational data describing the permanent dipole moment of ozone at vibrational equilibrium. With these simplifications, we can re-write $H_\mathrm{F}$ as

\begin{equation}
    H_\mathrm{F} = \frac{1}{2}\omega_\mathrm{c}^2q_\mathrm{c}^2 + \frac{1}{2}p_\mathrm{c}^2 + \frac{1}{2}\mu^2 \lambda_\mathrm{c}^2 - \omega_\mathrm{c} \mu \lambda_\mathrm{c} q_\mathrm{c}.
\end{equation}

The first two terms correspond to the classical total energy of the cavity mode, the third term is the dipole self-energy term more relevant in the ultrastrong coupling limit, and the last term, $H_\mathrm{LM}$, is the conventional light-matter interaction term on which we now focus our attention. To make further analytical progress, we assume that the molecule is close to its vibrational equilibrium state. By doing so, we can approximate the permanent dipole moment as $\mu \approx \mu_0 + \frac{\partial \mu}{\partial q} |_{q_0} (q-q_0)$ where $q$ is a generalized vibrational coordinate. Whether this assumption holds depends on the degree of vibrational excitation. For instance, this assumption encounters no issues with considering solely thermal population in the tri-atomic model, where the harmonic vibrational frequencies are several times higher than $k_\mathrm{B} T$. However, in our numerical studies, the molecules are initialized with energies larger than the dissociation energy that is itself several times larger than the vibrational frequencies, so throughout the IVR process, the vibrational mode displacements from equilibrium can be large. Nonetheless, for the purpose of providing a simple, analytic form of the Rabi splitting that can be generalized to higher orders if necessary, we plug this term into $H_\mathrm{LM}$:

\begin{equation}
    H_\mathrm{LM} = -\left[\omega_\mathrm{c}\mu_0 \lambda_\mathrm{c} q_\mathrm{c} + \frac{\partial \mu}{\partial q} \Big|_{q_0} (q-q_0) \omega_\mathrm{c}\lambda_\mathrm{c} q_\mathrm{c}\right].
\end{equation}

The latter term corresponds to the coupling between a vibrational transition and the cavity photon, and the Rabi splitting energy should correspond to the scaling factor of this term. We replace $q_\mathrm{c}$ and $q$ with quantized field operators:

\begin{equation}
    \frac{\partial \mu}{\partial q} \Big|_{q_0} \omega_\mathrm{c} q \lambda_\mathrm{c} q_\mathrm{c} =  \frac{\partial \mu}{\partial q} \Big|_{q_0} \omega_\mathrm{c} \sqrt{\frac{\hbar}{2M\omega_\mathrm{v}}}(\hat{b}^\dagger + \hat{b})\lambda_\mathrm{c} \sqrt{\frac{\hbar}{2\omega_\mathrm{c}}}(\hat{a}_\mathrm{c}^\dagger - \hat{a}_\mathrm{c}),
\end{equation}
where $M$ is the effective mass of the vibration, $\omega_\mathrm{v}$ is the frequency of the vibration, and $\hat{b}$ and $\hat{a}_\mathrm{c}$ are the annihilation operators of the vibrational and cavity modes, respectively.

We collect all the non-operator terms, multiply by a factor of two, and call this term the estimated Rabi splitting energy, or the energy difference between the lower and upper polariton:
\begin{equation}
    \hbar\Omega_\mathrm{R}^\mathrm{est} = \frac{\partial \mu}{\partial q} \Big|_{q_0} \hbar \lambda_\mathrm{c}\sqrt{\frac{\omega_\mathrm{c}}{M\omega_\mathrm{v}}}.
\end{equation}

In our numerical studies, we have found that the cavity mode influences the unimolecular dissociation rate most strongly when the cavity is resonant with a vibrational mode dominated by the local mode vibration, such as $q_3$ when the cavity is $x$-polarized, especially for vanishing local vibrational mode-mode momentum coupling $\epsilon$. Therefore, for the cavity field along the $x$-direction we approximate $\frac{\partial \mu}{\partial q} \big|_{q_0}$ as 

\begin{equation}
    \frac{\partial \mu}{\partial q} \Big|_{q_0}\approx\frac{\partial \mu}{\partial q_3} \Big|_{q_3^0}=-A\sin{(q_3^0/2)}(q_1^0+q_2^0)/2.
\end{equation}

\begin{figure}[!tbhp]
\centering
\includegraphics[width=\linewidth]{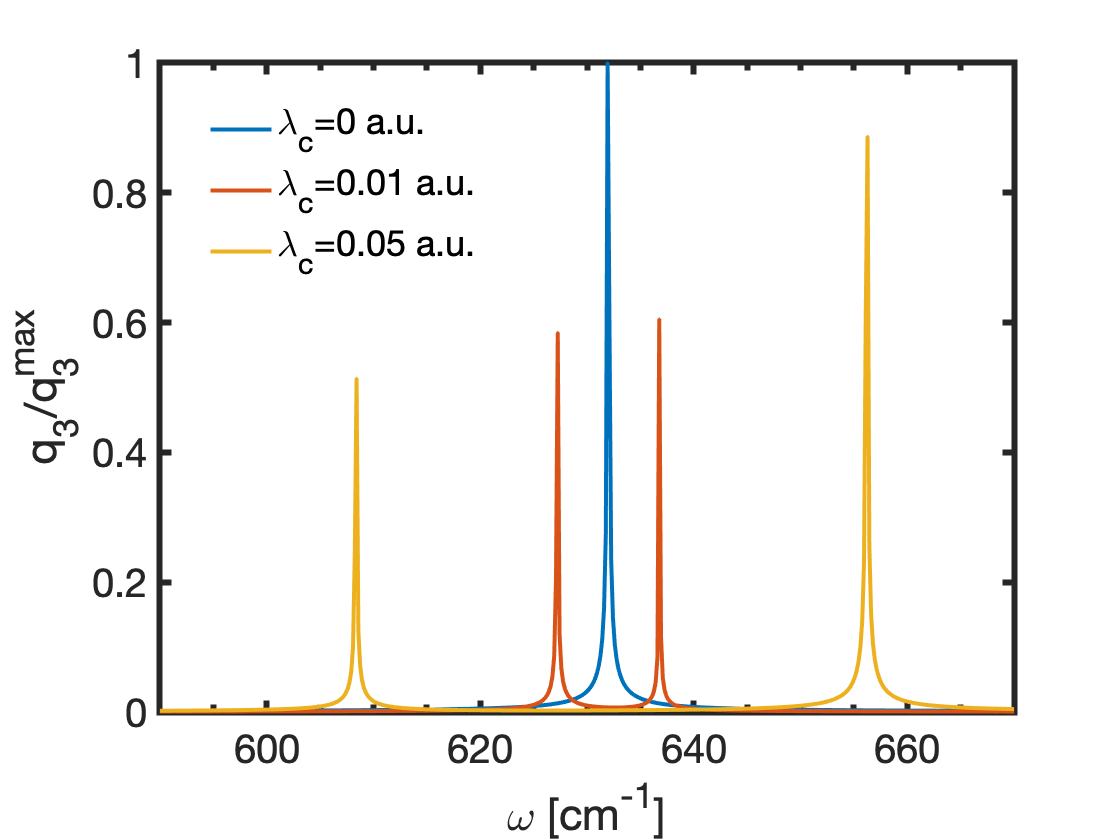}
\caption{Rabi splitting of the local bending mode $q_3$. We set $\epsilon=0$ and initialize the molecule with low enough energy such that the vibrational coordinates are close to equilibrium. The molecule is placed inside a cavity with frequency in resonance to the local bending mode, i.e. $\omega_{\mathrm c}=\omega_3$ for varying cavity strength $\lambda_{\mathrm c}$. The estimated Rabi splitting $\hbar\Omega_{\mathrm R}^\mathrm{est}\sim 10$ cm$^{-1}$ for $\lambda_{\mathrm c}=0.01$ a.u., closely matching the actual Rabi splitting observed.}
\label{fig:rabi}
\end{figure}

We estimate the Rabi splitting for typical parameters used in our numerical studies in atomic units, where $\hbar=1$. For instance, for $A=0.395$ a.u. so that the permanent dipole moment at equilibrium $\mu_0$ is 1 e$\cdot$Bohr, $q_1^0=q_2^0=2.416$ Bohr, $q_3^0=2.039$ rad, $M_{q_3} = 1/G_{33}^{(0)} = (2.88 \cdot 10^{-5})^{-1}$ a.u., $\omega_3=\omega_\mathrm{c}=2.88\cdot 10^{-3}$ a.u., and $\lambda_\mathrm{c}=0.01$ a.u., we estimate a Rabi splitting $\hbar\Omega_{\mathrm r}^\mathrm{est}=4.364\cdot 10^{-5}$ a.u. or $9.577$ cm$^{-1}$.
This estimate indeed matches the splittings observed in Fourier transforms of the vibrational coordinates of the coupled molecule-cavity system under the conditions described, as in Fig. \ref{fig:rabi}, where the molecule is initialized with energy $\ll \hbar\omega_i$ to prevent the appearance of overtones and combination spectra. We propagate the same initial condition for all $\lambda_c$: $\{p_1, p_2, p_3, q_1, q_2, q_3, p_\mathrm{c}, q_\mathrm{c} \} = \{5, 5, -5, 2.8, 2.8, 2, 0.01, 0.01\}$ in atomic units.

Experimental demonstrations of vibrational polariton chemistry rely on the collective coupling of many molecules to the cavity mode to achieve the Rabi splittings observed in the dipole moment spectra on the order of 10 cm$^{-1}$ to 100 cm$^{-1}$. It is thus desirable to extend our model that considers a single molecule coupled to the cavity to the collective case and explore the effects of disorder \cite{Du2021}, or ensemble-enhanced coupling of the cavity mode to a single molecule \cite{Schutz2020, Sidler2021}, among others. 

\section{Orthogonally polarized cavity} \label{app:ycavity}
\begin{figure}[!tbhp]
\centering
\includegraphics[width=1.0\linewidth]{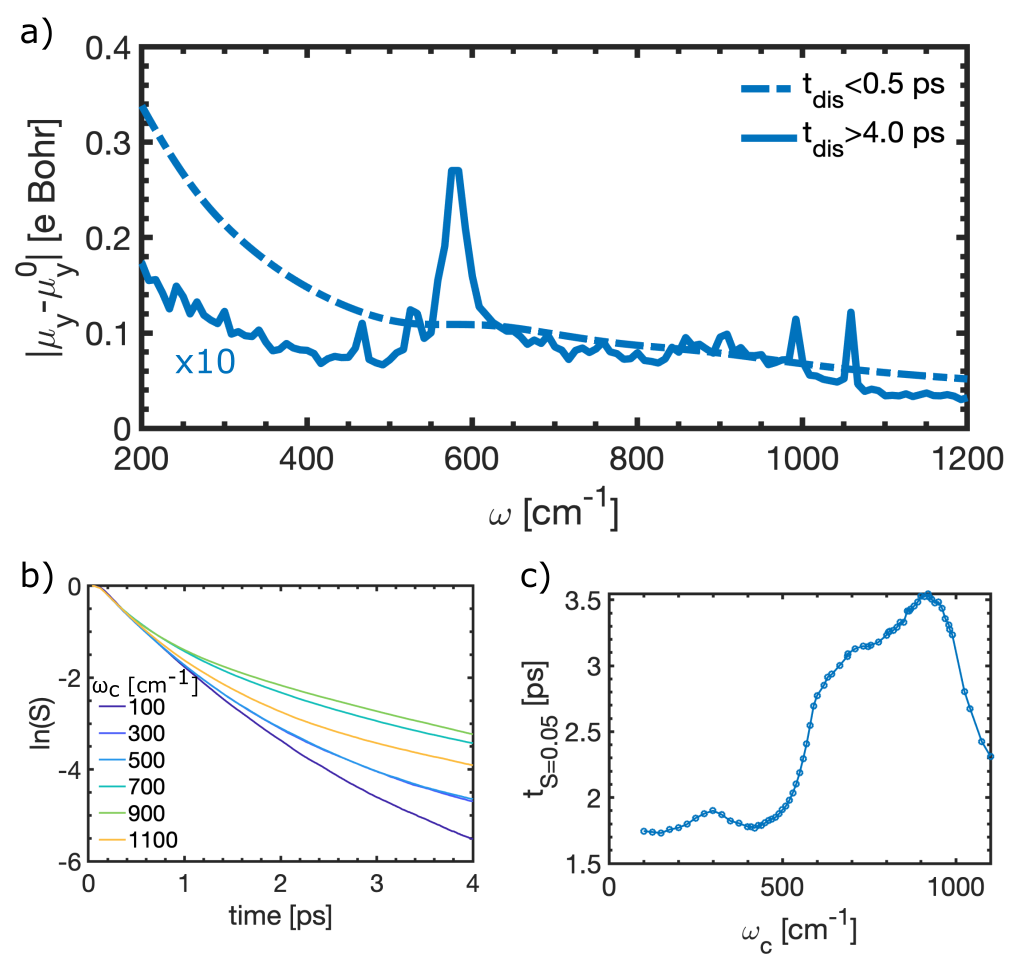}
\caption{\textbf{Dissociation in a $y$-polarized cavity. (a)} $y$-polarized dipole moment spectra $\mu_y(\omega)$ of short- and long-lived trajectories of the molecule outside the cavity. \textbf{(b)} Survival curves and \textbf{(c)} unimolecular dissociation rate for the model molecule in a $y$-polarized cavity. We vary cavity frequency $\omega_{\mathrm c}$ for cavity strength $\lambda_{\mathrm c}=0.05$ and observe a broadly resonant rate slow-down around $\sim$900 cm$^{-1}$.}
\label{fig:reactionrate_y}
\end{figure}

We also explore the effect of coupling the cavity to different molecular resonances. As we show in Fig. \ref{fig:reactionrate_y}(a), the $y$-polarized dipolar moment spectrum for long-lived trajectories has peaks around 600 and 1000 cm$^{-1}$, away from the resonances at 520 cm$^{-1}$ we explore in the main text. Therefore, we can expect the light-matter coupling interaction for a $y$-polarized cavity to also be relatively high at these frequencies. We set a constant cavity strength $\lambda_{\mathrm c}=0.05$ a.u. along the $y$-direction and then vary the cavity frequency $\omega_{\mathrm c}$ from 100 cm$^{-1}$ to 1100 cm$^{-1}$. Here, we observe a broader resonant effect at a different frequency of $\sim$900 cm$^{-1}$, corresponding to the broader peak in the $y$-polarized dipolar moment spectrum, as opposed to the sharper double peak $\sim$520 cm$^{-1}$ and $580$ cm$^{-1}$ in Fig. \ref{fig:reactionrate}(c)-(d) where the cavity field is oriented along the $x$-direction.

\FloatBarrier

\section{Small $\epsilon$} \label{sec:accelerate}

\begin{figure}[!tbhp]
\centering
\includegraphics[width=0.8\linewidth]{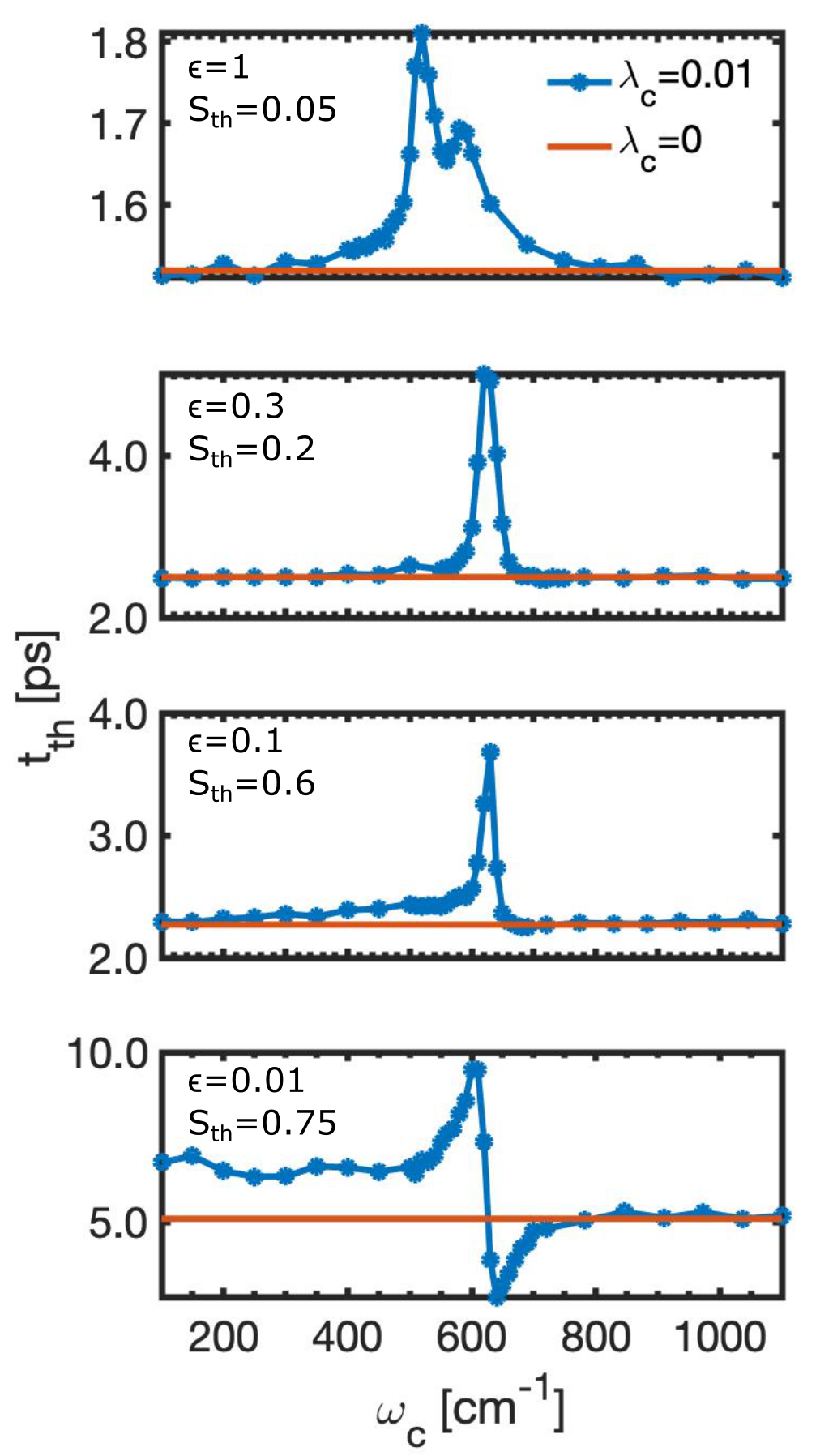}
\caption{Weak mode-mode coupling regime. Time $t_\mathrm{th}$ required for survival probability threshold $S_\mathrm{th}(t)=0.05$, 0.2, 0.6, and 0.75 as a function of cavity frequency  for vibrational mode-mode momentum coupling $\epsilon=1$, 0.3, 0.1, and 0.01, respectively. The blue line corresponds to the molecule inside the cavity with varying frequency $\omega_{\mathrm c}$ and cavity strength $\lambda_{\mathrm c}=0.01$, and the red line corresponds to the molecule outside the cavity, or $\lambda_{\mathrm c}=0$. At larger $\epsilon=1$, 0.3, and 0.1, the unimolecular dissociation rate only decreases inside a resonant cavity relative to outside the cavity, while at small $\epsilon=0.01$, the unimolecular dissociation rate can also be increased.
}
\label{fig:accelerate}
\end{figure}

To that end, we calculate the decay time of the initial reactants by varying $\epsilon \in \{1.0, 0.3, 0.1, 0.01\}$ with fixed cavity strength $\lambda_{\mathrm c}$ in Fig. \ref{fig:accelerate}. The survival probability thresholds $S_\mathrm{th} \in \{0.05, 0.2, 0.6, 0.7\}$ differ for each $\epsilon$ because as $\epsilon$ decreases, the timescale of decay increases, resulting in increased computational cost. Also, we note that lower $\epsilon$ roughly corresponds to \textit{e.g.} higher mass of the center atom relative to the other two atoms \cite{Bunker1962}. For $\epsilon=1.0, 0.3,$ and $0.1$, the time $t_\mathrm{th}$ to reach $S_\mathrm{th}$ inside the cavity is larger, or the reaction rate decreases, whereas for $\epsilon=0.01$, the time to $S(t)=0.75$ can also decrease at a particular resonance frequency. Hence, the reaction rate can be faster. This effective increase of the reaction rate can be interpreted as follows: for large $\epsilon$, the dynamics of energy distribution are dominated by vibrational mode-mode momentum coupling, whereas for small $\epsilon\ll 1$, the dynamics are dominated by coupling to the cavity, which then can assist in transferring energy between vibrational modes.

The case of relatively small $\epsilon$ can be loosely analogized to the case of \textit{intermolecular} vibrational energy redistribution, since the energies of intermolecular interactions are one to three orders of magnitude lower than intramolecular ones. Alternatively, the weak mode-mode momentum couplings could represent coupling to low-frequency intramolecular vibrational modes known to serve as baths for vibrational energy \cite{Uzer1991}. Regardless of the interpretation, the results in Fig. \ref{fig:accelerate} demonstrate that the cavity mode can effectively augment the coupling between vibrational modes and accelerate the reaction rate. These results may be relevant to the studies showing enhanced reaction rates when tuning the cavity into resonance with a vibrational mode of solvent molecules \cite{Lather2019, Xiang2020} that are vibrationally weakly coupled to reactant molecules outside the cavity. 

\FloatBarrier

\end{document}